\documentclass[oldversion]{aa}
\UseRawInputEncoding

\usepackage{graphicx}
\usepackage{txfonts}
\bibpunct{(}{)}{;}{a}{}{,} 

\usepackage{natbib}
\usepackage{relsize}

\def\sss{\scriptscriptstyle}
\def\U{{\sss \!\mathrm{U}}}
\def\L{{\sss \!\mathrm{L}}}
\def\K{{\sss \!\mathrm{K}}}

\def\S{{\sss \!S}}
\def\I{{\sss \!I}}
\def\C{{\sss \!C}}

\def\O{{\sss \!O}}
\def\R{{\sss \!R}}

\def\nur{\nu_\mathrm{r}}
\def\nuv{\nu_\theta}
\def\nuL{\nu_\L}
\def\nuU{\nu_\U}
\def\nuK{\nu_\K}

\def\RISCO{\R\I\S\C\O}

\def\RISCO{r_{\mathrm{\RISCO}}}

\begin{document}

\title{Models of high-frequency quasi-periodic oscillations and black hole spin estimates in Galactic microquasars}
\titlerunning{Models of QPOs and BH spin estimates in Galactic microquasars}


\author{
A.~Kotrlov\'{a}\inst{\ref{inst1}}\and
E.~\v{S}r\'{a}mkov\'{a}\inst{\ref{inst1}}\and
G.~T\"{o}r\"{o}k\inst{\ref{inst1}}\and
K.~Goluchov\'{a}\inst{\ref{inst1},\ref{inst5}}\and
J.~Hor\'{a}k\inst{\ref{inst2}}\and
O.~Straub\inst{\ref{inst3},\ref{inst4},\ref{inst1}}\and
D.~Lan\v{c}ov\'{a}\inst{\ref{inst1}}\and\\
Z.~Stuchl\'{\i}k\inst{\ref{inst5}}\and
M.~A.~Abramowicz\inst{\ref{inst1},\ref{inst6},\ref{inst7}}
}
\institute{Research Centre for Computational Physics and Data Processing, {Institute of Physics}, Silesian University in Opava, Bezru\v{c}ovo n\'am.~13, CZ-746\,01 Opava, Czech Republic \label{inst1}\\
\email{Andrea.Kotrlova@physics.slu.cz}
\and
Research Centre of Theoretical Physics and Astrophysics, Institute of Physics, Silesian University in Opava, Bezru\v{c}ovo n\'am.~13, CZ-746\,01 Opava, Czech Republic \label{inst5}
\and
Astronomical Institute, Academy of Sciences, Bo\v{c}n\'{\i} II 1401, CZ-14131 Praha 4-Spo\v{r}ilov, Czech Republic\label{inst2}
\and
Max Planck Institute for Extraterrestrial Physics, Gie\ss enbachstra\ss e 1, 85748 Garching, Germany\label{inst3}
\and
LESIA, Observatoire de Paris, Universit� PSL, CNRS,Sorbonne Universit�s, UPMC Univ.  Paris 06, Univ.  de Paris, Sorbonne Paris Cit�, 5 place Jules Janssen, 92195 Meudon, France\label{inst4}
\and
Nicolaus Copernicus Astronomical Centre, Polish Academy of Sciences, Bartycka 18, 00-716 Warsaw, Poland \label{inst6}
\and
Department of Physics, G\"{o}teborg University, Sweden \label{inst7}
}
\authorrunning{Kotrlov\'{a} et al.}

\date{Received / Accepted}

\abstract
{We explore the influence of non-geodesic pressure forces that are present in an accretion disk on the frequencies of its axisymmetric and non-axisymmetric epicyclic oscillation modes. {We discuss its implications for models of high-frequency quasi-periodic oscillations (QPOs) that have been observed in the X-ray flux of accreting black holes (BHs) in the three Galactic microquasars, GRS 1915+105, GRO J1655$-$40 and XTE J1550$-$564. We focus on previously considered QPO models that deal with low azimuthal number epicyclic modes, $\lvert m \rvert \leq 2$, and outline the consequences for the estimations of BH spin, $a\in[0,1]$.} For four out of six examined models, we find only small, rather insignificant changes compared to the geodesic case. For the other two models, on the other hand, there is a fair increase of the estimated upper limit on the spin. Regarding the QPO model's falsifiability, we find that one particular model from the examined set is incompatible with the data. If the microquasar's spectral spin estimates that point to $a>0.65$ were fully confirmed, two more QPO models would be ruled out. Moreover, if two very different values of the spin, such as $a\approx 0.65$ in GRO~J1655$-$40 vs. $a\approx 1$ in GRS~1915+105, were confirmed, all the models except one would remain unsupported by our results. Finally, we discuss the implications for a model recently proposed in the context of neutron star (NS) QPOs as a disk-oscillation-based modification of the relativistic precession model. This model provides overall better fits of the NS data and predicts more realistic values of the NS mass compared to the relativistic precession model. We conclude that it also implies a significantly higher upper limit on the microquasar's BH spin ($a\sim 0.75$ vs. $a\sim 0.55$).}


\maketitle


\section{Introduction}\label{Section:intro}


Studying the X-ray spectra and variability provides a powerful tool for putting constraints on properties of compact objects such as mass, $M$, and spin, $a\equiv\mathrm{c}J/(\mathrm{G}M^2)$, of a black hole (BH). Among promising methods to measure the BH spin is fitting the X-ray spectral continuum or the relativistically broadened iron K$\alpha$ lines \citep[][]{mcc-rem:2006,sha-etal:2008,ste-etal:2009}. Various approaches based on X-ray timing, which are complementary to spectral methods, have been gaining popularity as well. One of them is the determination of BH properties using the observations of high-frequency quasi-periodic oscillations (HF QPOs)\footnote{For the sake of simplicity, we often use the shorter term "QPOs" instead of "HF QPOs" throughout the paper.} and related proposed models.

Detections of elusive HF QPO peaks in Galactic microquasars are frequently reported at rather constant frequencies that usually appear in ratios of small natural numbers \citep{abr-klu:2001:,Rem-etal:2002:,mcc-rem:2006}. There often appear two peaks that form a 3:2 frequency ratio, $R=\nuU/\nuL= 3/2$, where $\nuU$ ($\nuL$) is the higher (lower) of the two QPO frequencies \citep[see, however,][]{bel-etal:2012,bel-alt:2013,var-rod:2018}.\footnote{The evidence for rational frequency ratios has also been discussed in the context of neutron star (NS) QPOs. In the NS sources, clustering of twin-peak QPO detections most frequently arises as a result of weakness of (one or both) QPOs outside the limited range of the QPO frequencies (frequency ratio), see \citet{abr-kar-etal:2003:}, \citet{bel-etal:2005}, \citet{bel-etal:2007}, \citet{tor-etal:2008b,tor-etal:2008a}, \citet{bar-bou:2008}, \citet{bou-etal:2010}.}

It has been argued that, since the QPO frequencies roughly correspond to time-scales of orbital motion in the vicinity of BHs, the phenomenon likely originates in the innermost parts of accretion disks or in their corona. A number of papers have been devoted to the discussion on the determination of $M$ and $a$ that stems from this premise. Different QPO models incorporate different physical concepts in which the QPO excitation radii are located within the most luminous accretion region, usually below $r=20\,r_{\mathrm{G}}$ (where $r_{\rm G}\equiv \mathrm{G}M/\mathrm{c}^2$). Several models, for example, assume that QPOs are produced by a local motion of accreted inhomogeneities, such as blobs or vortices. This subset of QPO models includes the so-called relativistic precession (RP) or the tidal disruption model \citep[][]{abr-etal:1992,ste-vie:1998,ste-vie:1999,cad-etal:2008,kos-etal:2009,bak-etal:2014,kar-etal:2017,ger-etal:2017}. Another possibility is to relate the QPOs to a collective motion of the accreted matter, in particular to some accretion disk oscillatory modes {that have been explored for both thin and thick disks \cite[][]{kat-fuk:1980:,oka-etal:1987,Now-Wag:1992,wag:1999,abr-klu:2001:,kat:2001:PASJ,wag-etal:2001,sil-etal:2001:,wan-etal:2015:MNRAS:,rez-etal:2003,tor-etal:2005,Ingram+Done:2010,fra-etal:2016,stu-etal:2012,stu-etal:2013,ort-etal:2020,stu-etal:2020:,Mas-etal:2020:}.} Despite the large efforts having been made over the past three decades to explain this phenomenon, and a good number of QPO models having been proposed, no clear consensus on what the precise physical mechanism responsible for its occurence might be has been reached until this point. In several massive extragalactic sources, features that are analogous to QPOs in microquasars (but occur at much lower frequencies) have also been reported and discussed, namely in the context of their central BH's properties \citep{gol-etal:2019:AAL, gup-etal:2019}.
%


The two decades ago proposed RP model is often used for the estimation of NS and BH parameters based on the QPOs. Miscellaneous other competing models have been utilized as well. It is well known, for instance, that the RP model predicts a rather low BH spin for Galactic microquasars, which is in contradiction with some spectral spin estimates. Numerous other estimates that are based on a large variety of QPO models have been carried out by various authors \citep[the list of references can be found in, e.g.,][]{tor-etal:2011:AA,gol-etal:2019:AAL}. Most of them have been carried out considering a geodesic accretion flow. In more general flows, non-geodesic effects connected to, e.g., pressure gradients, magnetic fields or other forces may have a potentially significant impact on the QPO based spin predictions.

In this work, we aim to quantify such impact in the particular case of a non-geodesic influence that originates in the pressure forces that are present in the accretion flow modeled by a slightly non-slender pressure-supported perfect fluid torus. We study a specific group of `disk-oscillation' models that involve various combinations of epicyclic modes of accretion disk oscillations. Following our recent work \citep[][]{sra-etal:2015,tor-etal:2016:MNRAS}, we primarily focus on comparing two prominent models. Firstly, a model that deals with axisymmetric modes that in the slender torus limit exhibit frequencies equal to the radial and vertical epicyclic frequencies of perturbed geodesic motion. Secondly, a model that in the slender torus limit exhibits the same observable frequencies as the RP model.


The paper is organized as follows. In Sections \ref{Section:tori} - \ref{Section:slender:torus}, we shortly recall the physical description of non-slender accretion disks, QPO models and the applied methodology. In Section~\ref{Section:RP0}, we focus on comparing BH spin estimates carried out based on the two above mentioned models. All relevant facts and results regarding these models are explored here in detail. Section~\ref{Section:other} provides a comprehensive extension of the approach introduced in Section~\ref{Section:RP0} to several other previously considered QPO models. In Section~\ref{Section:conclusions}, we provide the reader with a quick overall quantitative summary, and, furthermore, we explore the main consequences and state our main concluding remarks. {Finally, in the electronic Appendix, we provide a set of formulae determining frequencies of the considered epicyclic modes calculated for non-slender tori.}

\section{Accretion tori}
\label{Section:tori}

We consider oscillations of non-slender pressure-supported tori {(thick disks)} that are made by a perfect polytropic fluid and surround rotating Kerr BHs. Following our previous studies, we assume the specific angular momentum distribution of the flow (defined through the covariant time and azimuthal components of the flow velocity) to be constant within the whole volume of the torus,\footnote{We adopt the metric signature in the (-,+,+,+) form.}
\begin{equation}
\ell \equiv -\frac{u_\phi}{u_t} =\mathrm{const} = \ell_{\mathrm{c}},
\end{equation}
as oppose to the Keplerian distribution,
\begin{equation}
{\ell = \ell_{\mathrm{K}},}
\end{equation}
characteristic for geometrically thin Keplerian flows (thin disks), whose radial structure is mostly determined by a balance between the gravitational and inertial forces.


\begin{figure*}[ht]
\begin{center}
\includegraphics[width=0.9\hsize]{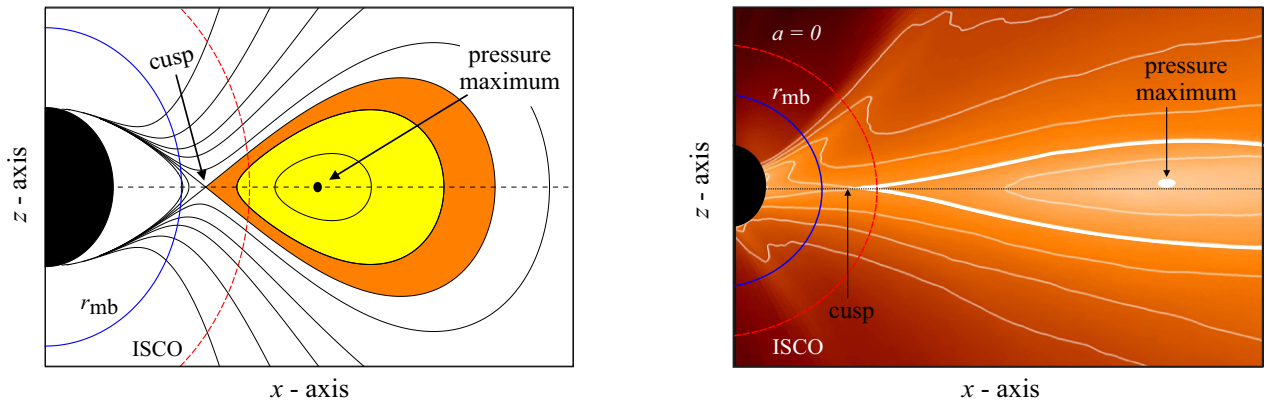}
\end{center}
\caption{\emph{Left:} Illustration of the topology of equipotential surfaces that determine the spatial distribution of fluid in thick disks (often called Polish doughnuts). The yellow region corresponds to tori of various thickness. The orange (along with yellow) region corresponds to a torus with a cusp. The topology allows for many disks with no cusp ($\beta < \beta_{\mathrm{cusp}}$) and one disk with a cusp ($\beta = \beta_{\mathrm{cusp}}$). The self-crossing equipotential curve corresponds to the marginally overflowing torus with $\ell_\K(r_{\mathrm{ms}}) < \ell_\mathrm{c}<\ell_\K(r_{\mathrm{mb}})$. The torus has a finite extent and is terminated by a cusp located at its inner edge. The coloured lines corresponding to constant radii denote the marginally stable (ISCO) and marginally bound ($r_{\mathrm{mb}}$) orbit. {A more detailed illustration along with rigorous classification of possible equipotential curves topologies can be found in \citet{Abr-Jar-Sik:1978:ASTRA:}.}  {\emph{Right:}} The equipressure contours seen within an up-to-date general relativistic three-dimensional global radiative magnetohydrodynamic simulation. The figure is based on the work of \citet{lan-etal:2019:ApJ} who have reported on a new class of realistic solutions of black hole accretion flows -- the so-called puffy accretion disks. {The setup of the simulation is very general and does not assume any form of initial toroidal structure of the fluid within the inner accretion region.}
\label{figure:1}}
\end{figure*}

\subsection{{Stable configurations of thick relativistic disks}}

For constant angular momentum tori, \citet{Abr-Jar-Sik:1978:ASTRA:} \citep[see also][]{koz-jar-abr:1978} have shown that equilibrium structures of the equipressure and equidensity surfaces coincide with those of the constant effective potential surfaces determined by
\begin{equation}
    W \equiv -u_t = \left(-g^{tt} + 2\ell_{\mathrm{c}} g^{t\phi} - \ell_{\mathrm{c}}^2
g^{\phi\phi}\right)^{-1/2} = \rm{const}.
    \label{eq:effective-potential}
\end{equation}
Here, $g^{\mu\nu}$ denote components of the spacetime metric expressed in the Boyer--Lindquist coordinates. The surface of the torus coincides with one of the equipotential surfaces where the pressure vanishes. The critical points (extrema and saddles) of the effective potential correspond to vanishing pressure gradients and thus to time-like circular geodesics. At these points, the rotation of the flow is Keplerian. The centre of the torus (i.e., the circle at $r=r_\mathrm{c}$ where the pressure has its maximum) corresponds to a stable time-like circular geodesic located at the local minimum of the effective potential in the equatorial plane.

{For angular momenta inside a specific range, there can be another unstable circular geodesic that corresponds to the saddle point of the effective potential and puts a limit on the possible size of the torus located at a given $r_\mathrm{c}$.} The related critical equipotential has a characteristic `cusp', through which the matter may be accreted onto the central BH without the need for any viscous processes, similarly as for the well-known Roche-lobe overflow in binary systems \citep{koz-jar-abr:1978}.

\subsection{{The inner edge}}

{While the marginally stable circular orbit, $r_\mathrm{ms}$, is often called the innermost stable circular orbit (ISCO) of a thin accretion disk and forms its inner edge, the inner edge of a thick disk has a different location. Situated between $r_\mathrm{ms}$ and the marginally bound circular orbit, $r_\mathrm{mb}$, this location depends on the disk's angular momentum, $\ell_{\mathrm{c}}$ \citep{Abr-Jar-Sik:1978:ASTRA:}. For $\ell_{\mathrm{c}} > \ell_\K(r_{\mathrm{mb}})$, the disk is infinite with its inner edge located at $r_{\mathrm{mb}}$, but still well inside the cusp self-crossing equipotential. The $\ell_{\mathrm{c}} = \ell_\K(r_{\mathrm{mb}})$ case corresponds to an infinite torus terminated by the cusp at $r_{\mathrm{mb}}$. For $\ell_{\mathrm{c}} < \ell_\K(r_{\mathrm{mb}})$, the configuration corresponds to a finite, marginally overflowing torus. This torus has its inner edge located closer to BH than $r_\mathrm{ms}$, but not closer than $r_\mathrm{mb}$.}

Examples of various tori configurations are illustrated in the left panel of Figure~\ref{figure:1}. While the equilibrium configuration of fluid that embodies the cusp equipotential represents a rather simplified stationary analytic model of a non-accreting disk, in accretion, the very existence of the cusp is of general importance. As long as the disk dynamical timescale is fairly shorter than the viscous timescale, which is true for most disk models, the cusp plays a key role in the accretion of matter onto the BH. Fluid elements above the cusp inside the torus stay at more or less the same radius for a large number of dynamical periods, whereas, below the cusp, accretion is more of a Bondi-like type. The presence of the cusp is indeed often seen in sophisticated numerical simulations of accretion flows \citep[e.g.][]{qia-etal:2009,fra-etal:2007,fra-etal:2009,gim-fon:2017,lan-etal:2019:ApJ}. This is illustrated in the right panel of Figure~\ref{figure:1}.

\subsection{Torus size}
\label{Section:size}

To quantify the torus size, \citet{abr-etal:2006} have introduced the `thickness' $\beta$ parameter,
\begin{equation}
    \beta =
\frac{\sqrt{2n}\,c_\mathrm{sc}}{{r_\mathrm{c}}\Omega_\mathrm{c} u^t_\mathrm{c}},
\end{equation}
where $n$ is the polytropic index and $c_\mathrm{sc}$, $u^t_\mathrm{c}$, $\Omega_\mathrm{c}$ are the polytropic sound speed, the contravariant time component of the four-velocity, and the angular velocity of the flow defined at the centre of the torus, $r=r_\mathrm{c}$. For angular momenta $\ell_\K(r_\mathrm{ms}) < \ell_{\mathrm{c}} < \ell_\K(r_\mathrm{mb})$, this parameter is limited by
\begin{equation}
    0\leq\beta\leq\beta_\mathrm{cusp} \equiv
\frac{\sqrt{2}}{r_\mathrm{c}\Omega_\mathrm{c} u^t_\mathrm{c}}\left(1 -
\frac{W_\mathrm{c}}{W_\mathrm{cusp}}\right)^{1/2},
\end{equation}
whereas, for $\ell_{\mathrm{c}} \geq \ell_\K(r_\mathrm{mb})$, the possible range of $\beta$ is given by
\begin{equation}
    0\leq\beta\leq\beta_\infty \equiv \frac{\sqrt{2}}{r_\mathrm{c}\Omega_\mathrm{c}
u^t_\mathrm{c}}\left(1 - W_\mathrm{c}\right)^{1/2}.
\end{equation}
Here, $W_\mathrm{c}$ and $W_\mathrm{cusp}$ are the effective potential values corresponding to the centre and cusp equipotential.

In general, the value of $\beta$ that corresponds to marginally overflowing tori, $\beta=\beta_\mathrm{cusp}(a)$, lies in the range
\begin{equation}
0\leq\beta_\mathrm{cusp}(a)\leq\beta_{\infty}(a),
\end{equation}
where $\beta_\mathrm{cusp}(a)=0$ describes a torus located at the marginally stable circular orbit, $r_{\mathrm{ms}}=r_\mathrm{ms}(a)$. For the purpose of the present work, it is useful to introduce the `effective' $\beta$ parameter,
\begin{equation}
\beta_\mathrm{eff}\equiv\beta/\beta_\infty.
\end{equation}
For $\ell_{\mathrm{c}}\geq\ell_\K(r_\mathrm{mb})$, the possible equilibrium configurations correspond to $0\leq \beta_\mathrm{eff}\leq 1$, while, for $\ell_\K(r_\mathrm{ms}) < \ell_{\mathrm{c}} <\ell_\K(r_\mathrm{mb})$, the allowed range is $0\leq\beta_\mathrm{eff}\leq \beta_\mathrm{cusp}/\beta_\infty$. Based on the value of $\beta_\mathrm{eff}$, one may determine whether a given cusp configuration corresponds to a small torus, or to a large torus with $\beta \approx \beta_\mathrm{\infty}$.

\section{QPO models under consideration}
\label{Section:models}

There is a large collection of papers suggesting that QPOs are related to oscillations of accretion tori \citep[][]{abr-klu:2001:,rez-etal:2003,abr-etal:2006,Ingram+Done:2010,fra-etal:2016,ave-etal:2017}. These studies often assume that oscillations of tori can be responsible for both BH and NS  QPOs, the large differences between the two classes of sources being related mostly to a different QPO modulation mechanism \citep[][]{bur-etal:2004:APJ,hor:2005,abr-etal:2006}. {A subset of these studies puts its largest focus on the epicyclic modes of torus oscillations. Based on the evidence for the 3:2 QPO frequency ratio, it is often speculated that the QPOs are connected to a nonlinear resonant coupling between different pairs of these modes.}

{In this work, we continue in the efforts to study the epicyclic modes of torus oscillations in the context of fitting the observed QPO frequencies.} We presume that the two oscillatory modes identified with the observed 3:2 QPOs are excited at the same radius and under the same physical conditions (i.e., the same torus configuration). This assumption is valid not only for the resonance based concepts, but also for a broader class of models. In this sense, our study is relevant to the consideration of the epicyclic oscillation modes in a more general context. {In next, as well as in our previous studies, we refer to different combinations of epicyclic modes as to different QPO models although they could be in principle  viewed as different versions of just one model, which deals with the epicyclic oscillations of thick disks.}

{Generally speaking, the probability of exciting a given mode with a certain amplitude decreases with increasing azimuthal wave number $m$. We consider situations characterized by $\lvert m \rvert \leq 2$. {We further restrict our attention to several models that are based on various physical motivations suggested in preceding studies.} In some cases, apart from disc oscillation modes, they also deal with the Keplerian circular motion. {Two of these models have been favoured since they exhibit the potential for resonant coupling (group A). The others represent alternatives of two models previously elaborated and supported within numerous papers in the context of thin disks (group B).} All these models except one have been utilized in the work of \citet{tor-etal:2011:AA} who calculated BH spin values for a purely geodesic flow in the three Galactic microquasars with HF QPOs -- GRS 1915+105, GRO J1655$-$40 and XTE J1550$-$564.}

{\subsection{Group A}}

No fully self-consistent concept of resonant models that would incorporate some of the epicyclic modes has been proposed so far. Despite it being a necessary requirement, the frequency commensurability is not a sufficient condition for the resonance to occur. Other important requirements follow from the symmetry properties of the involved oscillatory modes, such as their parities with respect to the equatorial plane or the azimuthal wavenumber. {Out of all oscillatory modes combinations discussed in this work, only the axisymmetric modes seem to fully satisfy these conditions \citep[][]{hor:2008}.}

{In this context, we examine the ``epicyclic'' (Ep) model of Abramowicz and Klu{\'z}niak, which attributes the HF QPOs to the axisymmetric radial and vertical epicyclic oscillation modes in the accretion disk. Alternatively, in the so-called Kep model, the two QPO frequencies are associated with the axisymmetric radial mode frequency and the Keplerian orbital frequency \cite[see][for details]{tor-etal:2005}. {Both models have been extensively studied by \citet{abr-klu:2001:,abr-kar-etal:2003:,klu-etal:2004,Hor-kar:2006,hor-etal:2009}.}}

{\subsection{Group B}}

{There are two combinations of modes that we denote as the ``RP1'' model \citep{bur:2005:} and the ``RP2'' model \citep{Tor-etal:2010}. In the slender torus limit, these two predict for a non-rotating BH the same observable frequencies as the RP model. In this limit, for any BH spin, the same observable frequencies as those predicted by the RP model are also predicted by another combination of modes that we in next denote as the RP0 model.} This model deals with the possibility of the observed QPO frequencies being associated to the $m=-1$ non-axisymmetric radial mode frequency and the Keplerian orbital frequency. We consider the RP0 model in addition to the set of models discussed by \citet{tor-etal:2011:AA}. The RP0 model is of special importance in the context of the recent findings on NS QPOs presented in the studies of \cite{tor-etal:2016:MNRAS,tor-etal:2018:MNRAS,tor-etal:2019}, which is further discussed in Section~\ref{Section:axi-non-axi}.

{Finally, we also assume} a combination of modes that in the slender torus limit leads to the same observable frequencies as the ``warped disk'' model proposed by \citet{kat:2001:PASJ,kat:2004:PASJ:tem} in the context of diskoseismology (the study of oscillations of thin disks). We denote it as the WD model.

\begin{figure*}
\begin{center}
\includegraphics[width=0.75\hsize]{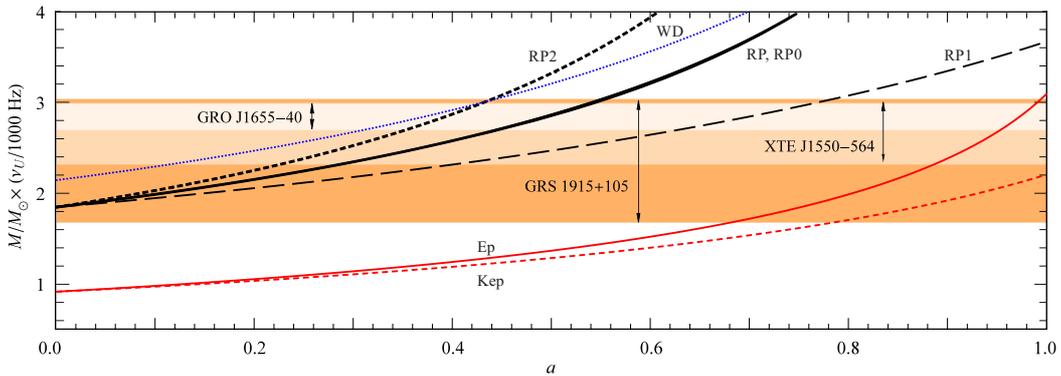}
\end{center}
\caption{The $M(a)$ curves implied by the geodesic QPO models. The RP0 model's predictions fully coincide with those of the RP model. The orange horizontal rectangles covering the full range of $a$ indicate the commonly accepted limits on the BH mass in each microquasar. \label{figure:F4}}
\end{figure*}

\section{Epicyclic mode frequencies}
\label{Section:slender:torus}

In Table~\ref{table:frequencies}, we recall the formulae for the observable QPO frequencies for each of the above models. In the slender torus limit, they are expressed in terms of frequencies of geodesic orbital motion, i.e., the Keplerian orbital frequency, $\nuK$, and the radial and vertical epicyclic frequency, $\nur$ and $\nuv$, that in the Boyer--Lindquist coordinates, $t,r,\theta,\phi$, may be written as \citep[e.g.,][]{ali-gal:1981,sil-etal:2001:,tor-stu:2005}
\begin{eqnarray}
\label{radial}
\nur^2 &=&\alpha_\mathrm{r}\,\nuK^2,
\\
\label{vertical}
\nu_{\theta}^2 &=&\alpha_\theta\,\nuK^2,
\end{eqnarray}
where
\begin{align}
\label{Keplerian}
&\nuK=\frac{1}{2\pi}\left({{\mathrm{G}M}\over {r_{\rm G}^{~3}}}\right)^{1/2}\left( x^{3/2} + a \right)^{-1},
\\
&\alpha_\mathrm{r}\left(x\,,a\right)\equiv{1-6\,x^{-1}+ 8 \,a \, x^{-3/2} -3 \, a^2 \, x^{-2}},
\\
&\alpha_\theta\left(x\,,a\right)\equiv{1-4\,a\,x^{-3/2}+3a^2\,x^{-2}},
\\
&x=r/r_{\rm G}, ~~~ r_{\rm G}=\mathrm{G}M/\mathrm{c}^2.\label{frekvence-def-konec}
\end{align}

\begin{table}[t]
\caption{
Frequency relations corresponding to QPO models considered in this work, listed for both the non-geodesic case and the slender torus limit.
\label{table:frequencies}}
\centering
\renewcommand{\arraystretch}{1.5}
\tabcolsep=6pt
\begin{tabular}{lcccc}
\hline
 & \multicolumn{4}{c}{Frequency relations}\\\cline{2-5}
Model & $\nuU$& $\nuU\,(\beta=0)$ & $\nuL$& $\nuL\,(\beta=0)$ \\
\hline
Ep  & $\nu^*_{\theta,0}$& $\nuv$            &  $\nu^*_{\mathrm{r},0}$& $\nur$ \\
Kep & $\nuK$ & $\nuK$                             & $\nu^*_{\mathrm{r},0}$& $\nur$ \\
\hline
RP0 & $\nuK$ & $\nuK$                              & $\nu^*_{\mathrm{r},-1}$& $\nuK-\nur$\\
RP1 & $\nu^*_{\theta,0}$&$\nuv$            & $\nu^*_{\mathrm{r},-1}$& $\nuK-\nur$ \\
RP2 &  $\nu^*_{\theta,-2}$& $2\nuK-\nuv$    & $\nu^*_{\mathrm{r},-1}$& $\nuK-\nur$\\
WD  & $\nu^*_{\mathrm{r},-2}$& $2\nuK-\nur$ & $2\nu^*_{\mathrm{r},-1}$& $2\left(\nuK-\nur\right)$\\
\hline
\end{tabular}
\end{table}

\subsection{Epicyclic mode frequencies in non-slender tori}
\label{Section:frequencies}

In non-slender tori, the frequencies of oscillation modes are modified by the pressure forces. As a result, the non-geodesic radial and vertical epicyclic frequencies will differ from those corresponding to the perturbed circular geodesic motion.

The study of oscillation and stability properties of fluid tori has been initiated by \citet{Pap-Pri:1984:MONNR:}, who have explored global linear stability of Newtonian fluid tori with respect to non-axisymmetric perturbations. Considering small linear perturbations to the torus equilibrium, they have derived a single partial differential equation governing the linear dynamics of oscillations of a Newtonian constant specific angular momentum torus \citep{Pap-Pri:1984:MONNR:}. Later on, general relativistic form of the Papaloizou--Pringle equation has been introduced by \citet{abr-etal:2006} and \citet{bla-etal:2006:}. In general, this equation cannot be fully solved analytically. Using a perturbation method, \citet{str-sram:2009} and {\citet{fra-etal:2016}} have derived fully general relativistic formulae determining the frequencies of axisymmetric and non-axisymmetric radial and vertical epicyclic modes in a slightly non-slender constant specific angular momentum torus within a second-order accuracy in the torus thickness.

Using relations~(\ref{radial}) -- (\ref{frekvence-def-konec}), the calculated frequencies of the radial and vertical epicyclic modes can be written in the following form:
\begin{eqnarray}
\label{frequencies-r}
\nu_{\mathrm{r},\,m}^*&=&\left[\sqrt{\alpha_\mathrm{r}} + m + \beta^2 C_{\mathrm{r},\,m}(r_{\rm{c}},a)\right]\nuK\\
&=&\phantom{mi}\nu_{\mathrm{r}} + \left[ m + \beta^2 C_{\mathrm{r},\,m}(r_{\rm{c}},a)\right]\nuK,\nonumber\\
\label{frequencies-t}
\nu_{\theta,\,m}^*&=&\left[\sqrt{\alpha_\theta} + m + \beta^2 C_{\theta,\,m}(r_{\rm{c}},a)\right]\nuK\\
&=&\phantom{mi}\nu_{\theta} + \left[m + \beta^2 C_{\theta,\,m}(r_{\rm{c}},a)\right]\nuK\nonumber.
\end{eqnarray}
Here, $m$ is the azimuthal wavenumber, and {$C_{\mathrm{r},\,m}(r_{\rm{c}},a)$ and $C_{\theta,\,m}(r_{\rm{c}},a)$} denote the negative second-order pressure corrections evaluated at the centre of the torus, $r=r_{\rm{c}}$. Since $C_{\mathrm{r},\,m}$ and $C_{\theta,\,m}$ are given by fairly long expressions, we provide their explicit form in the electronic Appendix.

\subsection{The approximative formulae applicability}

It is necessary to determine the range of the $\beta$ parameter relevant for our study. Formulae (\ref{frequencies-r}) and (\ref{frequencies-t}) should provide reasonable results for tori of moderate thickness, but they are not fully applicable for tori of a substantial width. To specify this statement quantitatively, a concrete physical situation needs to be taken into account. It is useful to express the torus thickness in terms of $\beta_\mathrm{eff}$. One can expect that $\beta_\mathrm{eff}\sim 0.3$ will likely yield a solid approximation of a real situation, while considering $\beta_\mathrm{eff}\sim 0.7$ could lead to incorrect results. In next, we use $\beta_{\mathrm{eff}}=0.3$ and $\beta_{\mathrm{eff}}=0.7$ as the referential values.

\section{BH spin estimation -- Ep and RP0 model}\label{Section:RP0}

In order to obtain the constraints on $M$ and $a$, we make a comparison between the expected and the observed QPO frequencies. Following \citet{tor-etal:2011:AA}, we compare the predicted frequencies to the observed frequencies in the specific case of the 3:2 frequency ratio (see Section~\ref{Section:conclusions} for a further discussion). The predictions of geodesic QPO models are illustrated in Figure~\ref{figure:F4}. \citet{sra-etal:2015} have applied the results of \citet{str-sram:2009} to study the particular case of the Ep model. Following the previous studies on BH spin estimations \citep[][]{klu-abr:2001,tor-etal:2005,tor-etal:2011:AA}, they have utilized the QPO independent mass estimates. The main conclusion of their work is that the effect of the pressure forces on the predicted QPO frequencies is very small when $a<0.9$. The influence becomes significant only for rapidly rotating BHs ($a>0.9$).

\citet{fra-etal:2016} have derived corrections to formulae for non-geodesic epicyclic frequencies assuming the exact form of the relativistic Papaloizou--Pringle equation as oppose to the approximative form considered by \citet{bla-etal:2006:,bla-etal:2007} and \citet{str-sram:2009}. We here use their formulae to revise the calculations of \citet{sra-etal:2015} carried out for the Ep model, and, furthermore, to extend this approach to another QPO model -- the RP0 model.

\subsection{The QPO frequencies behaviour}

\begin{figure}[t]
\begin{center}
\includegraphics[width=0.85\hsize]{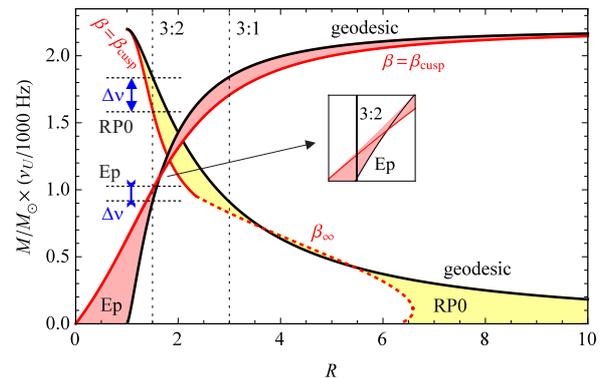}
\end{center}
\caption{
The upper QPO frequency predicted by the Ep and RP0 model plotted for $a=0$ and tori whose thickness ranges from an infinitely slender torus ($\beta=0$, black line) through a torus with a cusp ($\beta=\beta_{\mathrm{cusp}}$, red line) to a torus whose outer edge extends to infinity ($\beta_{\mathrm{eff}} = 1$, dotted red line). The blue arrows indicate the spread of the resonant frequency implied by the allowed spread of $\beta$ for each model and the $3\!:\!2$ QPO frequency ratio. 
\label{figure:F7a}}
\end{figure}

\begin{figure*}[t]
\begin{center}
\includegraphics[width=0.9\hsize]{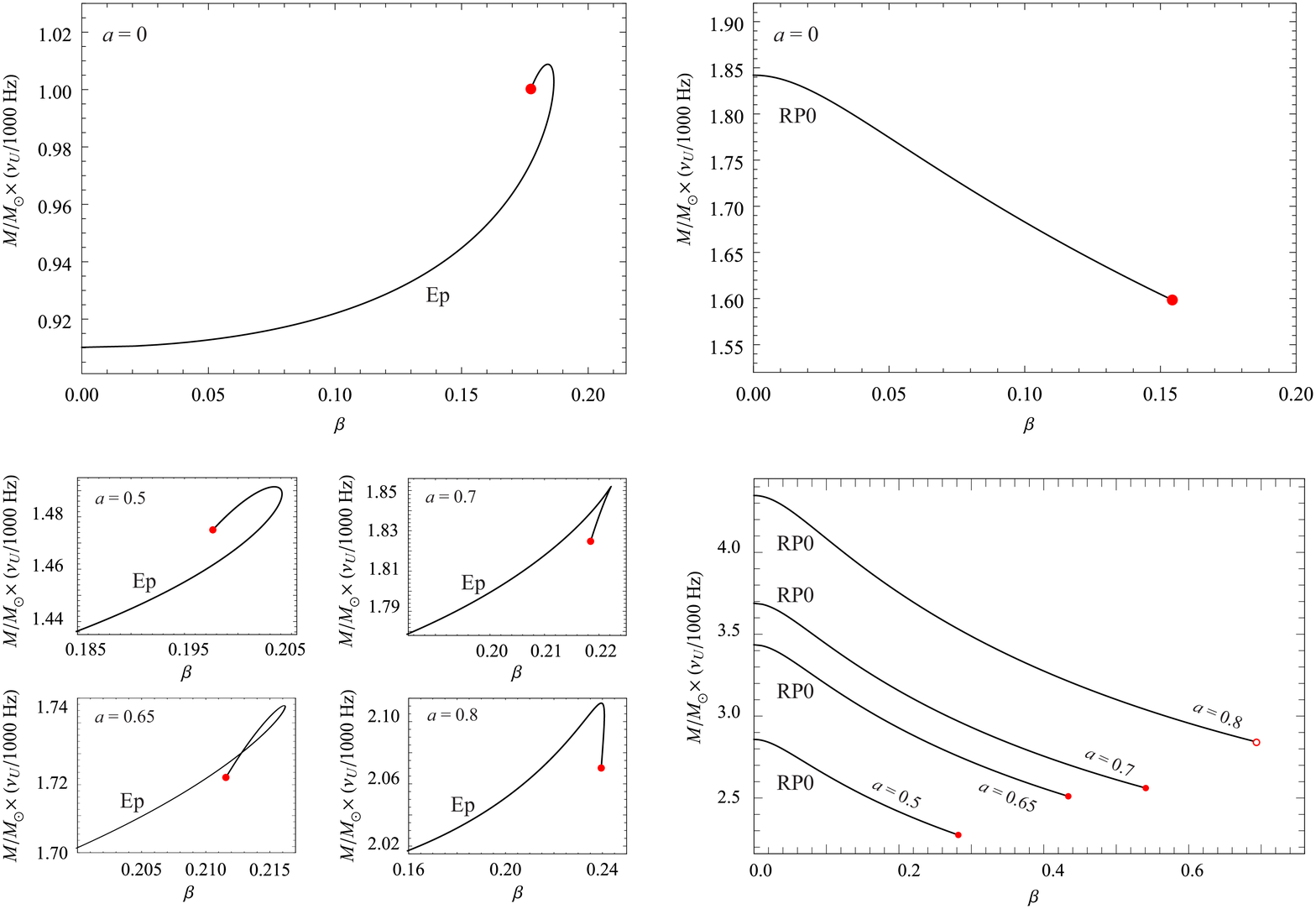}
\end{center}
\caption
{{Comparison between the upper QPO frequencies predicted by the Ep and RP0 models. \emph{Left:} Examples of the non-trivial topology of  $\nuU(\beta)$  curves predicted by the Ep model. Except for the $a=0$ case, we do not display the whole effective range of $\beta\in[0,\,\beta_{\mathrm{cusp}}]$ - the overall increase of $\nuU$ is rather small for any $a\lesssim 0.9$. \emph{Right:} The monotonic behaviour of $\nuU(\beta)$ functions predicted by the RP0 model. The values of $a$ are the same as in the left panel. In both panels, the red dots correspond to $\beta=\beta_{\mathrm{cusp}}$. The empty red circle denotes $\beta=\beta_{\infty}$.}
\label{compressedA}}
\end{figure*}

Although we mostly focus on the 3:2 frequency ratio, other frequency ratios are explored as well. {Figure~\ref{figure:F7a}} shows the results of such investigation for the Schwarzschild spacetime and tori whose thickness ranges from an infinitely slender torus to a torus with its outer edge extending to infinity. The QPO frequency ratio in this Figure is not fixed and takes values of up to $R=10$.

The enlarged area in {Figure~\ref{figure:F7a}} illustrates the QPO frequency behaviour relevant for the Ep model close to $R =3:2$. It is apparent that for ratios from this area, the QPO frequency's extremal value does not always correspond to a torus with a cusp. Such phenomenon has been discussed in more detail in our previous paper~\citep{sra-etal:2015}.

The position of the orbit that gives the 3:2 QPO frequency ratio within the Ep model changes in a non-trivial way as the torus size increases. As a consequence, the relation between $\beta$ and the QPO frequency is not always a function \citep[][]{bla-etal:2006:,sra-etal:2015} -- see the loops on the Ep model curves illustrated in {Figure~\ref{compressedA}}. Nevertheless, for $a=0$, these frequencies only display a small variation across the whole range of the allowed torus thickness. The small sensitivity of the predicted QPO frequency towards the torus thickness persists up to $a\sim 0.9$.

For the RP0 model and the 3:2 frequency ratio, the predicted QPO frequency is a monotonic function of $\beta$. {This is illustrated in Figure~\ref{compressedA} that shows a comparison between the Ep and RP0 models. Note that, for $a\in[0,\,0.9]$, the RP0 model is associated with a much higher quantitative impact of the torus thickness on the QPO frequency.}

\begin{figure}[t]
\begin{center}
\includegraphics[width=0.9\hsize]{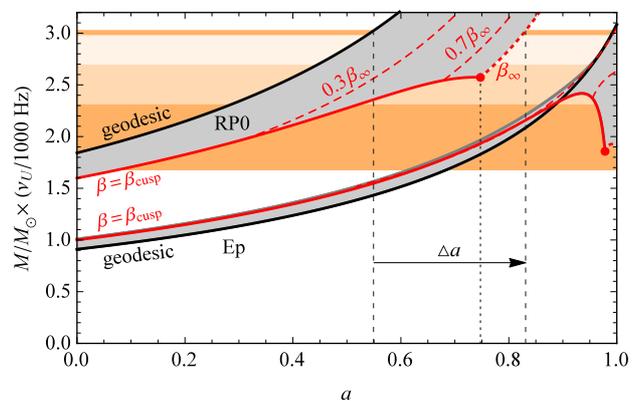}
\end{center}
\caption{
The $M(a)$ relation implied by the RP0 and Ep model. The thick black curves correspond to the geodesic case, i.e., these are the same curves as those shown in Figure~\ref{figure:F4}. The gray shaded region indicates the case when $\beta>0$. The thick red curves correspond to $\beta=\beta_{\mathrm{cusp}}$ and the dotted red curve to $\beta = \beta_{\infty}$. The dashed red lines correspond to $\beta_{\mathrm{eff}} = \beta/\beta_{\infty} = 0.7$ and $\beta_{\mathrm{eff}} = \beta/\beta_{\infty} = 0.3$. The black arrow labelled $\Delta a$ indicates the shift of the upper limit on the spin of microquasars implied by the~RP0 model considering the~non-geodesic flow (from $a\sim0.55$ to $a\sim0.83$).
\label{figure:F7b}}
\end{figure}

\subsection{The main implications for BH spin}

\begin{figure*}
\begin{center}
\includegraphics[width=.9\hsize]{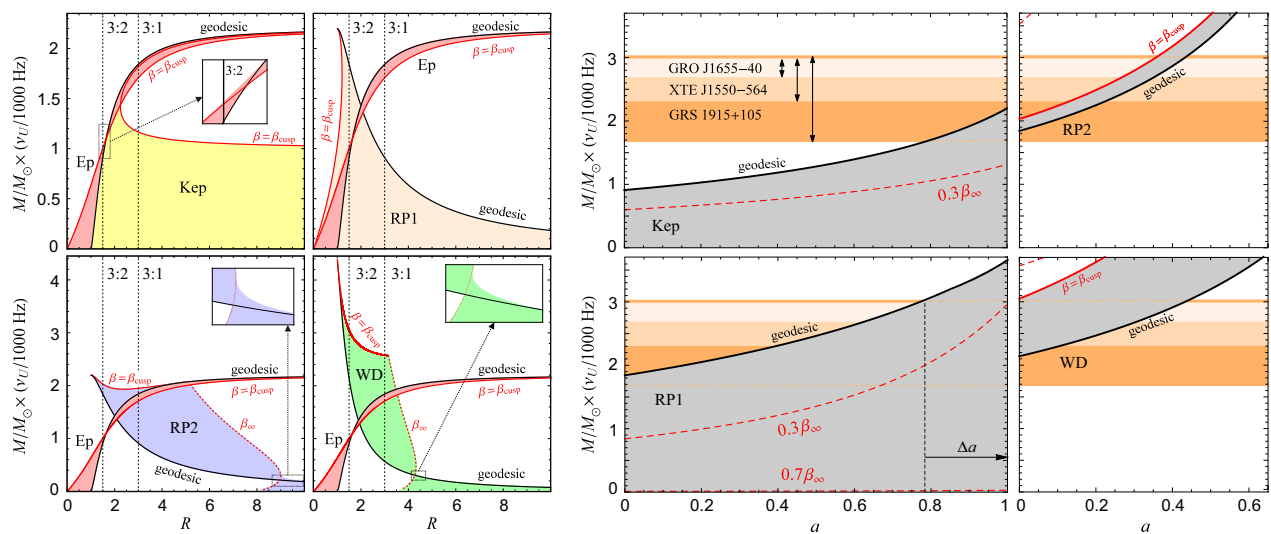}
\end{center}
\caption{
{\emph{Left: }}
The same as in {Figure~\ref{figure:F7a}}, but for the Kep, RP1, RP2 and WD model.
{\emph{Right: }}
The same as in {Figure~\ref{figure:F7b}}, but for the Kep, RP1, RP2 and WD model. Note that for the RP2 and WD model, we have $\beta \ll 0.3\beta_{\infty}$.
\label{figure:F8}}
\end{figure*}

{Figure~\ref{figure:F7b}} shows a comparison between the overall behaviour of the QPO frequencies implied by the Ep and RP0 model as it changes with increasing BH spin and torus thickness. Different curves in the Figure indicate the referential values of the relative torus thickness, $\beta_{\mathrm{eff}}\in {0, 0.3, 0.7, 1}$. There is only a small spread of the resonant frequency predicted by the Ep model for any but a very high spin ($a\gtrsim 0.9$). The lower limit on the spin given by this model, $a\approx 0.69$ for $\beta_{\mathrm{eff}}=0$, reaches the value of $a\approx0.62$ for the maximal allowed torus thickness. The impact of the non-geodesic effects is clearly more important within the RP0 model. For a geodesic limit of this model (which corresponds to the RP model), the maximal value of the spin is about $a\sim0.55$. Regarding larger tori, it can grow up to $a\sim0.75$ for tori with a cusp, or even to $a\sim0.83$ when tori with no cusp are assumed. A detailed quantification of the impact for each microquasar is presented in Table~\ref{table:spin}.

\section{BH spin estimation -- other models}
\label{Section:other}

We plot the spread of the upper QPO frequency implied by the Kep, RP1, RP2 and WD model for $a=0$ in the left panel of Figure~\ref{figure:F8}. The Ep model's predictions are presented as well, for the sake of comparison. In analogy to {Figure~\ref{figure:F7a}}, the QPO frequency ratio in this Figure is not fixed and takes values of up to $R=10$. For the 3:2 frequency ratio, the spread of the QPO frequency is rather large within the Kep, RP1, and WD model (contrary to the Ep and RP2 model). {We find that similar behaviour of the QPO model frequencies arises also for rotating BHs.}

The right panel of Figure~\ref{figure:F8} illustrates the impact of the above described QPO frequencies behaviour on the estimation of BH spin. The QPO frequencies in the limit of $\beta=0$ are increasing monotonic functions of $a$. The pressure corrections to these frequencies are negative for the Kep and RP1 model, while they are positive for the WD and RP2 model. Out of these four models, only for the RP1 model, there is a significant change of the estimated overall upper limit on the spin.

For the Kep model, there is a large spread of the predicted QPO frequency, but the scaled frequency $M\times\nuU(a)$ is clearly lower than the observationally constrained values for any $a\lesssim 0.79$. For this reason, the overall limit on the spin remains unchanged. Within the RP1 model, the geodesic curve enters most of the observationally constrained region $(a\in[0,~0.78])$, and the influence of a non-zero torus thickness raises the upper limit to $a=1$. Within the WD and RP2 model, the scaled frequencies $M\times\nuU(a)$ are higher than the observationally constrained values for any $a\gtrsim 0.44$. The estimated overall upper limit therefore remains unaltered for both these models.


\section{Discussion and conclusions}\label{Section:conclusions}

Table~\ref{table:spin} indicates how the limit on the spin changes for different models in each microquasar. In all the three microquasars, the main conclusions remain mostly unaltered compared to the geodesic case for four out of the six examined models. In particular, there is a small change of the lower limit on the spin in all the three sources in the Ep model's case. For the Kep, WD and RP2 model, the spin estimates for GRS 1915+105 remain fully unaltered, while for the other two sources, there is only a certain decrease of the lower limit on the inferred spin. On the other hand, within the RP0 and RP1 model, the estimated upper limit on the BH spin shows a significant increase.

\subsection{The main implications}

Regarding the QPO models' falsifiability, we conclude that much like in the geodesic case the Kep model is fully incompatible with the GRO J1655$-$40 and XTE 1550$-$564 data. Provided that the mechanism responsible for the QPO phenomenon is the same in all the three microquasars, this model would be ruled out. Although there is currently no final agreement on the spectral spin estimates, the overall sample of the iron line and continuum based studies suggests that at least one of the microquasars should exceed the value of $a=0.65$ \citep[][]{MCC-etal:2006,Mid-etal:2006,blu-etal:2009,mcc-etal:2014,mil:2015}. If this was confirmed, our study would in addition to the Kep model rule out also the WD and RP2 model. Moreover, if two very different values of the spin, such as $a\approx 0.65$ in GRO~J1655$-$40 vs. $a\approx 1$ in GRS~1915+105, were confirmed, all the models except the RP1 model would remain unsupported by our results.

\subsection{{Favoured models}}\label{Section:axi-non-axi}

As briefly mentioned in Sections \ref{Section:intro}--\ref{Section:models}, two of the models considered in this work are of special importance. The first one is the Ep model, which is prominent within the class of the resonance models proposed by \citet{abr-klu:2001:} and \citet{tor-etal:2005}. This model, which involves the axisymmetric modes, is by far the most developed resonance model of QPOs. Using the solution of the improved relativistic Papaloizou--Pringle equation, we confirm the previous result of \citet{sra-etal:2015} that indicates small sensitivity of the resonant frequency to the torus thickness and the requirement of a high BH spin. The second one is the RP0 model, which provides outstanding results regarding the overall context of matching the BH and NS HF QPOs.

{In the slender torus limit, the RP0 model predicts the same QPO frequencies as the RP model. For the marginally overflowing torus ($\beta=\beta_{\mathrm{cusp}}$), it merges with a model recently discussed by \cite{tor-etal:2016:MNRAS} in the context of NS QPOs -- hereafter the CT model. This model has been suggested as the RP model alternative which deals with torus oscillations. It is based on the expectation that cusp configurations are likely to appear in real accretion flows, in which case the actual overall accretion rate through the inner edge of the disk can be strongly modulated by the disk oscillations \citep{pac-1982,abr-etal:2006}.  The CT model provides generally better fits of the NS data than the RP model. It also predicts a lower NS mass than the RP model which, in some cases, implies a mass estimate that is too high \citep{tor-etal:2018:MNRAS,tor-etal:2019}.}

{The spin predicted by the CT model is given by the upper solid red curve in Figure~\ref{figure:F7b} (The RP0 model and $\beta=\beta_{\mathrm{cusp}}$). We conclude that the upper limit on the spin implied by this model is significantly higher than in the RP model's case, namely $a\sim0.75$ vs. $a\sim 0.55$. This is presumably in better agreement with the spectral spin estimates.}

\begin{table*}[t]
\caption{Intervals of spin implied for the~three microquasars by the considered QPO models for the geodesic ($a$) and non-geodesic ($a^*$) case.}
\label{table:spin}
\centering
\renewcommand{\arraystretch}{1.4}
\tabcolsep=6pt
\begin{tabular}{l cc c cc c cc}
\hline
 & \multicolumn{2}{c}{GRS~1915$+$105} && \multicolumn{2}{c}{XTE~J1550$-$564} && \multicolumn{2}{c}{GRO~J1655$-$40}\\
\cline{2-3} \cline{5-6} \cline{8-9}
Model & $a\sim$ & $a^*\sim$ && $a\sim$ & $a^*\sim$ && $a\sim$ & $a^*\sim$\\
\hline
  Ep  & $0.69 - 0.99$ & $0.62 - 1$  && $0.89 - 0.99$ & $0.86 - 1$ && $0.96 - 0.99$ &  $0.95 - 1$ \\
  Kep & $0.79 - 1$ & $0.79 - 1$  && $-$ & $-$ && $-$ &  $-$ \\
\hline
  RP0 & $<0.55$ & $<0.83$  && $0.29 - 0.54$ & $0.29 - 0.82$ && $0.45 - 0.53$ &  $0.45 - 0.82$ \\
  RP1 & $<0.78$ & $0 - 1$  && $0.41 - 0.76$ & $0.41 - 1$ && $0.63 - 0.76$ &  $0.63 - 1$ \\
  RP2 & $< 0.44$ & $<0.44$  && $0.23 - 0.43$ & $0.13 - 0.43$ && $0.36 - 0.43$ &  $0.27 - 0.43$ \\
	WD  & $<0.44$ & $<0.44$  && $0.12 - 0.43$ & $<0.43$ && $0.31 - 0.42$ &  $<0.42$ \\
\hline
\end{tabular}
\end{table*}

\subsection{Caveats}\label{caveats}

One should be aware of the limitations associated with the adopted perturbative approach. The results of our calculations that stem from the consideration of $\beta_{\mathrm{eff}}\gtrsim 0.7$ should be confronted with exact numerical treatment of the Papaloizou--Pringle equation.\footnote{The relation between $\beta$ and $\beta_{\mathrm{eff}}$ depends on the specific torus configuration underlying a given QPO  model. For all tori configurations considered in this paper, we have $\beta=0.3\beta_{\infty}\approx 0.2$ and $\beta=0.7\beta_{\infty}\in [0.4,\,0.6]\approx 0.5$.} A similar reservation applies to the consideration of $a\gg 0.9$, since, for high spins, the epicyclic mode frequencies are very sensitive to even small changes of the torus thickness.

It is not fully clear to what degree our assumptions match the real situation. We compare the observed 3:2 QPO frequencies with frequencies of the oscillation modes calculated for one particular torus configuration. There are not many observations of HF QPOs available for BH binaries, and even fewer of those that are available display the two peaks simultaneously. The integration time required for the QPO identification is a few orders of magnitude longer than the characteristic QPO period. Furthermore, there are observations suggesting that the BH HF QPO frequencies may vary in time and form continuous correlations similar to those observed in NSs that reach into (but are not necessarily constrained to) the 3:2 frequency ratio range \citep[][]{bel-etal:2012,bel-alt:2013,mot-bel-ste:2014,var-rod:2018}. Despite these uncertainties, our assumptions are sufficient for the presented simplified analysis that provides a brief comparison of BH spin estimates associated to models that deal with different combinations of disk oscillation epicyclic modes.

\begin{acknowledgements}
We acknowledge the Czech Science Foundation (GA\v{C}R) grant No.~17-16287S. We also wish to thank the INTER-EXCELLENCE project No.~LTI17018 that supports the collaboration between the Silesian University in Opava and the Astronomical Institute in Prague. Furthermore, we acknowledge two internal grants of the Silesian University, SGS/12,13/2019. Last but not least, we would like to express our thanks to the referee whose valuable comments and suggestions have greatly helped to improve the paper.
\end{acknowledgements}

\bibliographystyle{aa} 
\bibliography{reference-AA}

\begin{thebibliography}{79}
\expandafter\ifx\csname natexlab\endcsname\relax\def\natexlab#1{#1}\fi

\bibitem[{{Abramowicz} {et~al.}(1978){Abramowicz}, {Jaroszynski}, \&
  {Sikora}}]{Abr-Jar-Sik:1978:ASTRA:}
{Abramowicz}, M., {Jaroszynski}, M., \& {Sikora}, M. 1978, A\&A, 63, 221

\bibitem[{{Abramowicz} {et~al.}(2006){Abramowicz}, {Blaes}, {Hor{\'a}k},
  {Klu{\'z}niak}, \& {Rebusco}}]{abr-etal:2006}
{Abramowicz}, M.~A., {Blaes}, O.~M., {Hor{\'a}k}, J., {Klu{\'z}niak}, W., \&
  {Rebusco}, P. 2006, Classical and Quantum Gravity, 23, 1689

\bibitem[{{Abramowicz} {et~al.}(2003){Abramowicz}, {Karas}, {Klu{\'z}niak},
  {Lee}, \& {Rebusco}}]{abr-kar-etal:2003:}
{Abramowicz}, M.~A., {Karas}, V., {Klu{\'z}niak}, W., {Lee}, W.~H., \&
  {Rebusco}, P. 2003, PASJ, 55, 467

\bibitem[{{Abramowicz} \& {Klu{\'z}niak}(2001)}]{abr-klu:2001:}
{Abramowicz}, M.~A. \& {Klu{\'z}niak}, W. 2001, A\&A, 374, L19

\bibitem[{{Abramowicz} {et~al.}(1992){Abramowicz}, {Lanza}, {Spiegel}, \&
  {Szuszkiewicz}}]{abr-etal:1992}
{Abramowicz}, M.~A., {Lanza}, A., {Spiegel}, E.~A., \& {Szuszkiewicz}, E. 1992,
  Nature, 356, 41

\bibitem[{{Aliev} \& {Galtsov}(1981)}]{ali-gal:1981}
{Aliev}, A.~N. \& {Galtsov}, D.~V. 1981, Gen. Relativ. and Gravitation, 13, 899

\bibitem[{{Bakala} {et~al.}(2014){Bakala}, {T{\"o}r{\"o}k}, {Karas}, {Dov{\v
  c}iak}, {Wildner}, {Wzientek}, {{\v S}r{\'a}mkov{\'a}}, {Abramowicz},
  {Goluchov{\'a}}, {Mazur}, \& {Vincent}}]{bak-etal:2014}
{Bakala}, P., {T{\"o}r{\"o}k}, G., {Karas}, V., {et~al.} 2014, MNRAS, 439, 1933

\bibitem[{{Barret} \& {Boutelier}(2008)}]{bar-bou:2008}
{Barret}, D. \& {Boutelier}, M. 2008, New Astron. Rev., 51, 835

\bibitem[{{Belloni} {et~al.}(2005){Belloni}, {M{\'e}ndez}, \&
  {Homan}}]{bel-etal:2005}
{Belloni}, T., {M{\'e}ndez}, M., \& {Homan}, J. 2005, A\&A, 437, 209

\bibitem[{{Belloni} {et~al.}(2007){Belloni}, {M{\'e}ndez}, \&
  {Homan}}]{bel-etal:2007}
{Belloni}, T., {M{\'e}ndez}, M., \& {Homan}, J. 2007, MNRAS, 376, 1133

\bibitem[{{Belloni} \& {Altamirano}(2013)}]{bel-alt:2013}
{Belloni}, T.~M. \& {Altamirano}, D. 2013, MNRAS, 432, 10

\bibitem[{{Belloni} {et~al.}(2012){Belloni}, {Sanna}, \&
  {M{\'e}ndez}}]{bel-etal:2012}
{Belloni}, T.~M., {Sanna}, A., \& {M{\'e}ndez}, M. 2012, MNRAS, 426, 1701

\bibitem[{{Blaes} {et~al.}(2006){Blaes}, {Arras}, \&
  {Fragile}}]{bla-etal:2006:}
{Blaes}, O.~M., {Arras}, P., \& {Fragile}, P.~C. 2006, \mnras, 369, 1235

\bibitem[{{Blaes} {et~al.}(2007){Blaes}, {{\v S}r{\'a}mkov{\'a}}, {Abramowicz},
  {Klu{\'z}niak}, \& {Torkelsson}}]{bla-etal:2007}
{Blaes}, O.~M., {{\v S}r{\'a}mkov{\'a}}, E., {Abramowicz}, M.~A.,
  {Klu{\'z}niak}, W., \& {Torkelsson}, U. 2007, \apj, 665, 642

\bibitem[{{Blum} {et~al.}(2009){Blum}, {Miller}, {Fabian}, {Miller}, {Homan},
  {van der Klis}, {Cackett}, \& {Reis}}]{blu-etal:2009}
{Blum}, J.~L., {Miller}, J.~M., {Fabian}, A.~C., {et~al.} 2009, APJ, 706, 60

\bibitem[{{Boutelier} {et~al.}(2010){Boutelier}, {Barret}, {Lin}, \&
  {T{\"o}r{\"o}k}}]{bou-etal:2010}
{Boutelier}, M., {Barret}, D., {Lin}, Y., \& {T{\"o}r{\"o}k}, G. 2010, MNRAS,
  401, 1290

\bibitem[{{Bursa}(2005)}]{bur:2005:}
{Bursa}, M. 2005, in RAGtime 6/7: Workshops on black holes and neutron stars,
  ed. S.~{Hled{\'{\i}}k} \& Z.~{Stuchl{\'{\i}}k}, 39--45

\bibitem[{{Bursa} {et~al.}(2004){Bursa}, {Abramowicz}, {Karas}, \&
  {Klu{\'z}niak}}]{bur-etal:2004:APJ}
{Bursa}, M., {Abramowicz}, M.~A., {Karas}, V., \& {Klu{\'z}niak}, W. 2004,
  APJL, 617, L45

\bibitem[{{de Avellar} {et~al.}(2018){de Avellar}, {Porth}, {Younsi}, \&
  {Rezzolla}}]{ave-etal:2017}
{de Avellar}, M.~G.~B., {Porth}, O., {Younsi}, Z., \& {Rezzolla}, L. 2018,
  MNRAS, 474, 3967

\bibitem[{{Fragile} {et~al.}(2007){Fragile}, {Blaes}, {Anninos}, \&
  {Salmonson}}]{fra-etal:2007}
{Fragile}, P.~C., {Blaes}, O.~M., {Anninos}, P., \& {Salmonson}, J.~D. 2007,
  APJ, 668, 417

\bibitem[{{Fragile} {et~al.}(2009){Fragile}, {Lindner}, {Anninos}, \&
  {Salmonson}}]{fra-etal:2009}
{Fragile}, P.~C., {Lindner}, C.~C., {Anninos}, P., \& {Salmonson}, J.~D. 2009,
  APJ, 691, 482

\bibitem[{{Fragile} {et~al.}(2016){Fragile}, {Straub}, \&
  {Blaes}}]{fra-etal:2016}
{Fragile}, P.~C., {Straub}, O., \& {Blaes}, O. 2016, MNRAS, 461, 1356

\bibitem[{{German{\`a}}(2017)}]{ger-etal:2017}
{German{\`a}}, C. 2017, PRD, 96, 103015

\bibitem[{{Gimeno-Soler} \& {Font}(2017)}]{gim-fon:2017}
{Gimeno-Soler}, S. \& {Font}, J.~A. 2017, A\&A, 607, A68

\bibitem[{{Goluchov{\'a}} {et~al.}(2019){Goluchov{\'a}}, {T{\"o}r{\"o}k}, {{\v
  S}r{\'a}mkov{\'a}}, {Abramowicz}, {Stuchl{\'{\i}}k}, \&
  {Hor{\'a}k}}]{gol-etal:2019:AAL}
{Goluchov{\'a}}, K., {T{\"o}r{\"o}k}, G., {{\v S}r{\'a}mkov{\'a}}, E., {et~al.}
  2019, \aap, 622, L8

\bibitem[{{Gupta} {et~al.}(2019){Gupta}, {Tripathi}, {Wiita}, {Kushwaha},
  {Zhang}, \& {Bambi}}]{gup-etal:2019}
{Gupta}, A.~C., {Tripathi}, A., {Wiita}, P.~J., {et~al.} 2019, MNRAS, 484, 5785

\bibitem[{{Hor{\'a}k}(2005)}]{hor:2005}
{Hor{\'a}k}, J. 2005, Astronomische Nachrichten, 326, 845

\bibitem[{{Hor{\'a}k}(2008)}]{hor:2008}
{Hor{\'a}k}, J. 2008, A\&A, 486, 1

\bibitem[{{Hor{\'a}k} {et~al.}(2009){Hor{\'a}k}, {Abramowicz}, {Klu{\'z}niak},
  {Rebusco}, \& {T{\"o}r{\"o}k}}]{hor-etal:2009}
{Hor{\'a}k}, J., {Abramowicz}, M.~A., {Klu{\'z}niak}, W., {Rebusco}, P., \&
  {T{\"o}r{\"o}k}, G. 2009, A\&A, 499, 535

\bibitem[{{Hor{\'a}k} \& {Karas}(2006)}]{Hor-kar:2006}
{Hor{\'a}k}, J. \& {Karas}, V. 2006, A\&A, 451, 377

\bibitem[{{Ingram} \& {Done}(2010)}]{Ingram+Done:2010}
{Ingram}, A. \& {Done}, C. 2010, MNRAS, 405, 2447

\bibitem[{{Karssen} {et~al.}(2017){Karssen}, {Bursa}, {Eckart}, {Valencia-S},
  {Dov{\v c}iak}, {Karas}, \& {Hor{\'a}k}}]{kar-etal:2017}
{Karssen}, G.~D., {Bursa}, M., {Eckart}, A., {et~al.} 2017, MNRAS, 472, 4422

\bibitem[{{Kato}(2001)}]{kat:2001:PASJ}
{Kato}, S. 2001, PASJ, 53, 1

\bibitem[{{Kato}(2004)}]{kat:2004:PASJ:tem}
{Kato}, S. 2004, PASJ, 56, 905

\bibitem[{{Kato} \& {Fukue}(1980)}]{kat-fuk:1980:}
{Kato}, S. \& {Fukue}, J. 1980, APJ, 32, 377

\bibitem[{{Klu{\'z}niak} \& {Abramowicz}(2001)}]{klu-abr:2001}
{Klu{\'z}niak}, W. \& {Abramowicz}, M.~A. 2001, Acta Phys. Polonica B, 32, 3605

\bibitem[{{Klu{\'z}niak} {et~al.}(2004){Klu{\'z}niak}, {Abramowicz}, {Kato},
  {Lee}, \& {Stergioulas}}]{klu-etal:2004}
{Klu{\'z}niak}, W., {Abramowicz}, M.~A., {Kato}, S., {Lee}, W.~H., \&
  {Stergioulas}, N. 2004, AJ, 603, L89

\bibitem[{{Kosti{\'c}} {et~al.}(2009){Kosti{\'c}}, {{\v C}ade{\v z}},
  {Calvani}, \& {Gomboc}}]{kos-etal:2009}
{Kosti{\'c}}, U., {{\v C}ade{\v z}}, A., {Calvani}, M., \& {Gomboc}, A. 2009,
  A\&A, 496, 307

\bibitem[{{Kozlowski} {et~al.}(1978){Kozlowski}, {Jaroszynski}, \&
  {Abramowicz}}]{koz-jar-abr:1978}
{Kozlowski}, M., {Jaroszynski}, M., \& {Abramowicz}, M.~A. 1978, A\&A, 63, 209

\bibitem[{{Lan{\v{c}}ov{\'a}} {et~al.}(2019){Lan{\v{c}}ov{\'a}}, {Abarca},
  {Klu{\'z}niak}, {Wielgus}, {Saḑowski}, {Narayan}, {Schee}, {T{\"o}r{\"o}k},
  \& {Abramowicz}}]{lan-etal:2019:ApJ}
{Lan{\v{c}}ov{\'a}}, D., {Abarca}, D., {Klu{\'z}niak}, W., {et~al.} 2019, APJ,
  884, L37

\bibitem[{{Maselli} {et~al.}(2020){Maselli}, {Pappas}, {Pani}, {Gualtieri},
  {Motta}, {Ferrari}, \& {Stella}}]{Mas-etal:2020:}
{Maselli}, A., {Pappas}, G., {Pani}, P., {et~al.} 2020, APJ, 899, 139

\bibitem[{{McClintock} {et~al.}(2014){McClintock}, {Narayan}, \&
  {Steiner}}]{mcc-etal:2014}
{McClintock}, J.~E., {Narayan}, R., \& {Steiner}, J.~F. 2014, \ssr, 183, 295

\bibitem[{{McClintock} \& {Remillard}(2006)}]{mcc-rem:2006}
{McClintock}, J.~E. \& {Remillard}, R.~A. 2006, {Black hole binaries}
  (Cambridge University Press), 157--213

\bibitem[{{McClintock} {et~al.}(2006){McClintock}, {Shafee}, {Narayan},
  {Remillard}, {Davis}, \& {Li}}]{MCC-etal:2006}
{McClintock}, J.~E., {Shafee}, R., {Narayan}, R., {et~al.} 2006, APJ, 652, 518

\bibitem[{{Middleton} {et~al.}(2006){Middleton}, {Done}, {Gierli{\'n}ski}, \&
  {Davis}}]{Mid-etal:2006}
{Middleton}, M., {Done}, C., {Gierli{\'n}ski}, M., \& {Davis}, S.~W. 2006,
  MNRAS, 373, 1004

\bibitem[{{Miller} \& {Miller}(2015)}]{mil:2015}
{Miller}, M.~C. \& {Miller}, J.~M. 2015, \physrep, 548, 1

\bibitem[{{Motta} {et~al.}(2014){Motta}, {Belloni}, {Stella},
  {Mu{\~n}oz-Darias}, \& {Fender}}]{mot-bel-ste:2014}
{Motta}, S.~E., {Belloni}, T.~M., {Stella}, L., {Mu{\~n}oz-Darias}, T., \&
  {Fender}, R. 2014, MNRAS, 437, 2554

\bibitem[{{Nowak} \& {Wagoner}(1992)}]{Now-Wag:1992}
{Nowak}, M.~A. \& {Wagoner}, R.~V. 1992, APJ, 393, 697

\bibitem[{{Okazaki} {et~al.}(1987){Okazaki}, {Kato}, \&
  {Fukue}}]{oka-etal:1987}
{Okazaki}, A.~T., {Kato}, S., \& {Fukue}, J. 1987, PASJ, 39, 457

\bibitem[{{Ortega-Rodr{\'\i}guez} {et~al.}(2020){Ortega-Rodr{\'\i}guez},
  {Sol{\'\i}s-S{\'a}nchez}, {{\'A}lvarez-Garc{\'\i}a}, \&
  {Dodero-Rojas}}]{ort-etal:2020}
{Ortega-Rodr{\'\i}guez}, M., {Sol{\'\i}s-S{\'a}nchez}, H.,
  {{\'A}lvarez-Garc{\'\i}a}, L., \& {Dodero-Rojas}, E. 2020, MNRAS, 492, 1755

\bibitem[{{Paczynski} \& {Abramowicz}(1982)}]{pac-1982}
{Paczynski}, B. \& {Abramowicz}, M.~A. 1982, APJ, 253, 897

\bibitem[{{Papaloizou} \& {Pringle}(1984)}]{Pap-Pri:1984:MONNR:}
{Papaloizou}, J.~C.~B. \& {Pringle}, J.~E. 1984, \mnras, 208, 721

\bibitem[{{Qian} {et~al.}(2009){Qian}, {Abramowicz}, {Fragile}, {Hor{\'a}k},
  {Machida}, \& {Straub}}]{qia-etal:2009}
{Qian}, L., {Abramowicz}, M.~A., {Fragile}, P.~C., {et~al.} 2009, A\&A, 498,
  471

\bibitem[{{Remillard} {et~al.}(2002){Remillard}, {Muno}, {McClintock}, \&
  {Orosz}}]{Rem-etal:2002:}
{Remillard}, R.~A., {Muno}, M.~P., {McClintock}, J.~E., \& {Orosz}, J.~A. 2002,
  APJ, 580, 1030

\bibitem[{{Rezzolla} {et~al.}(2003){Rezzolla}, {Yoshida}, \&
  {Zanotti}}]{rez-etal:2003}
{Rezzolla}, L., {Yoshida}, S., \& {Zanotti}, O. 2003, MNRAS, 344, 978

\bibitem[{{Shafee} {et~al.}(2008){Shafee}, {McKinney}, {Narayan},
  {Tchekhovskoy}, {Gammie}, \& {McClintock}}]{sha-etal:2008}
{Shafee}, R., {McKinney}, J.~C., {Narayan}, R., {et~al.} 2008, APJL, 687, L25

\bibitem[{{Silbergleit} {et~al.}(2001){Silbergleit}, {Wagoner}, \&
  {Ortega-Rodr{\'{\i}}guez}}]{sil-etal:2001:}
{Silbergleit}, A.~S., {Wagoner}, R.~V., \& {Ortega-Rodr{\'{\i}}guez}, M. 2001,
  APJ, 548, 335

\bibitem[{{Steiner} {et~al.}(2009){Steiner}, {McClintock}, {Remillard},
  {Narayan}, \& {Gou}}]{ste-etal:2009}
{Steiner}, J.~F., {McClintock}, J.~E., {Remillard}, R.~A., {Narayan}, R., \&
  {Gou}, L. 2009, APJL, 701, L83

\bibitem[{{Stella} \& {Vietri}(1998)}]{ste-vie:1998}
{Stella}, L. \& {Vietri}, M. 1998, APJL, 492, L59

\bibitem[{{Stella} \& {Vietri}(1999)}]{ste-vie:1999}
{Stella}, L. \& {Vietri}, M. 1999, Phys. Rev. Lett., 82, 17

\bibitem[{{Straub} \& {{\v S}r{\'a}mkov{\'a}}(2009)}]{str-sram:2009}
{Straub}, O. \& {{\v S}r{\'a}mkov{\'a}}, E. 2009, Classical and Quantum
  Gravity, 26, 055011

\bibitem[{{Stuchl{\'\i}k} {et~al.}(2020){Stuchl{\'\i}k}, {Kolo{\v{s}}},
  {Kov{\'a}{\v{r}}}, {Slan{\'y}}, \& {Tursunov}}]{stu-etal:2020:}
{Stuchl{\'\i}k}, Z., {Kolo{\v{s}}}, M., {Kov{\'a}{\v{r}}}, J., {Slan{\'y}}, P.,
  \& {Tursunov}, A. 2020, Universe, 6, 26

\bibitem[{{Stuchl{\'\i}k} {et~al.}(2012){Stuchl{\'\i}k}, {Kotrlov{\'a}}, \&
  {T{\"o}r{\"o}k}}]{stu-etal:2012}
{Stuchl{\'\i}k}, Z., {Kotrlov{\'a}}, A., \& {T{\"o}r{\"o}k}, G. 2012, \actaa,
  62, 389

\bibitem[{{Stuchl{\'\i}k} {et~al.}(2013){Stuchl{\'\i}k}, {Kotrlov{\'a}}, \&
  {T{\"o}r{\"o}k}}]{stu-etal:2013}
{Stuchl{\'\i}k}, Z., {Kotrlov{\'a}}, A., \& {T{\"o}r{\"o}k}, G. 2013, \aap,
  552, A10

\bibitem[{{T{\"o}r{\"o}k} {et~al.}(2008{\natexlab{a}}){T{\"o}r{\"o}k},
  {Abramowicz}, {Bakala}, {Bursa}, {Hor{\'a}k}, {Klu{\'z}niak}, {Rebusco}, \&
  {Stuchlik}}]{tor-etal:2008b}
{T{\"o}r{\"o}k}, G., {Abramowicz}, M.~A., {Bakala}, P., {et~al.}
  2008{\natexlab{a}}, Acta Astron., 58, 15

\bibitem[{{T{\"o}r{\"o}k} {et~al.}(2005){T{\"o}r{\"o}k}, {Abramowicz},
  {Klu{\'z}niak}, \& {Stuchl{\'{\i}}k}}]{tor-etal:2005}
{T{\"o}r{\"o}k}, G., {Abramowicz}, M.~A., {Klu{\'z}niak}, W., \&
  {Stuchl{\'{\i}}k}, Z. 2005, \aap, 436, 1

\bibitem[{{T{\"o}r{\"o}k} {et~al.}(2008{\natexlab{b}}){T{\"o}r{\"o}k},
  {Bakala}, {Stuchlik}, \& {\v{C}ech}}]{tor-etal:2008a}
{T{\"o}r{\"o}k}, G., {Bakala}, P., {Stuchlik}, Z., \& {\v{C}ech}, P.
  2008{\natexlab{b}}, Acta Astron., 58, 1

\bibitem[{{T{\"o}r{\"o}k} {et~al.}(2010){T{\"o}r{\"o}k}, {Bakala}, {{\v
  S}r{\'a}mkov{\'a}}, {Stuchl{\'{\i}}k}, \& {Urbanec}}]{Tor-etal:2010}
{T{\"o}r{\"o}k}, G., {Bakala}, P., {{\v S}r{\'a}mkov{\'a}}, E.,
  {Stuchl{\'{\i}}k}, Z., \& {Urbanec}, M. 2010, \apj, 714, 748

\bibitem[{{T{\"o}r{\"o}k} {et~al.}(2016){T{\"o}r{\"o}k}, {Goluchov{\'a}},
  {Hor{\'a}k}, {{\v S}r{\'a}mkov{\'a}}, {Urbanec}, {Pech{\'a}{\v c}ek}, \&
  {Bakala}}]{tor-etal:2016:MNRAS}
{T{\"o}r{\"o}k}, G., {Goluchov{\'a}}, K., {Hor{\'a}k}, J., {et~al.} 2016,
  MNRAS, 457, L19

\bibitem[{{T{\"o}r{\"o}k} {et~al.}(2018){T{\"o}r{\"o}k}, {Goluchov{\'a}}, {{\v
  S}r{\'a}mkov{\'a}}, {Hor{\'a}k}, {Bakala}, \&
  {Urbanec}}]{tor-etal:2018:MNRAS}
{T{\"o}r{\"o}k}, G., {Goluchov{\'a}}, K., {{\v S}r{\'a}mkov{\'a}}, E., {et~al.}
  2018, MNRAS, 473, L136

\bibitem[{{T{\"o}r{\"o}k} {et~al.}(2019){T{\"o}r{\"o}k}, {Goluchov{\'a}},
  {{\v{S}}r{\'a}mkov{\'a}}, {Urbanec}, \& {Straub}}]{tor-etal:2019}
{T{\"o}r{\"o}k}, G., {Goluchov{\'a}}, K., {{\v{S}}r{\'a}mkov{\'a}}, E.,
  {Urbanec}, M., \& {Straub}, O. 2019, MNRAS, 488, 3896

\bibitem[{{T{\"o}r{\"o}k} {et~al.}(2011){T{\"o}r{\"o}k}, {Kotrlov{\'a}}, {{\v
  S}r{\'a}mkov{\'a}}, \& {Stuchl{\'{\i}}k}}]{tor-etal:2011:AA}
{T{\"o}r{\"o}k}, G., {Kotrlov{\'a}}, A., {{\v S}r{\'a}mkov{\'a}}, E., \&
  {Stuchl{\'{\i}}k}, Z. 2011, A\&A, 531, A59

\bibitem[{{T{\"o}r{\"o}k} \& {Stuchl{\'{\i}}k}(2005)}]{tor-stu:2005}
{T{\"o}r{\"o}k}, G. \& {Stuchl{\'{\i}}k}, Z. 2005, A\&A, 437, 775

\bibitem[{{{\v C}ade{\v z}} {et~al.}(2008){{\v C}ade{\v z}}, {Calvani}, \&
  {Kosti{\'c}}}]{cad-etal:2008}
{{\v C}ade{\v z}}, A., {Calvani}, M., \& {Kosti{\'c}}, U. 2008, A\&A, 487, 527

\bibitem[{{{\v S}r{\'a}mkov{\'a}} {et~al.}(2015){{\v S}r{\'a}mkov{\'a}},
  {T{\"o}r{\"o}k}, {Kotrlov{\'a}}, {Bakala}, {Abramowicz}, {Stuchl{\'{\i}}k},
  {Goluchov{\'a}}, \& {Klu{\'z}niak}}]{sra-etal:2015}
{{\v S}r{\'a}mkov{\'a}}, E., {T{\"o}r{\"o}k}, G., {Kotrlov{\'a}}, A., {et~al.}
  2015, \aap, 578, A90

\bibitem[{{Varniere} \& {Rodriguez}(2018)}]{var-rod:2018}
{Varniere}, P. \& {Rodriguez}, J. 2018, APJ, 865, 113

\bibitem[{{Wagoner}(1999)}]{wag:1999}
{Wagoner}, R.~V. 1999, Physics Reports, 311, 259

\bibitem[{{Wagoner} {et~al.}(2001){Wagoner}, {Silbergleit}, \&
  {Ortega-Rodr{\'{\i}}guez}}]{wag-etal:2001}
{Wagoner}, R.~V., {Silbergleit}, A.~S., \& {Ortega-Rodr{\'{\i}}guez}, M. 2001,
  AJ, 559, L25

\bibitem[{{Wang} {et~al.}(2015){Wang}, {Chen}, {Zhang}, {Lei}, {Qu}, \&
  {Song}}]{wan-etal:2015:MNRAS:}
{Wang}, D.~H., {Chen}, L., {Zhang}, C.~M., {et~al.} 2015, MNRAS, 454, 1231

\end{thebibliography}

\end{document}